\documentclass[letterpaper,10pt,conference]{ieeeconf}  
\IEEEoverridecommandlockouts                              
\usepackage{cite} 
\usepackage{epsf} 
\usepackage{graphicx} 
\usepackage{color}
\usepackage{amssymb} 
\usepackage{wrapfig} 
\usepackage{mathtools} 
\usepackage{subfigure,bbm} 
\usepackage{amsmath} 
\usepackage{color} 
\usepackage{subcaption}  
\usepackage[font=footnotesize]{caption}  
\usepackage[colorlinks,linkcolor=blue,anchorcolor=blue,citecolor=blue]{hyperref} 
\usepackage{algorithmic} 
\usepackage{array} 
\usepackage{cancel} 
\usepackage{graphicx} 
\usepackage{color}
\usepackage{amssymb} 
\usepackage{wrapfig} 
\usepackage{mathtools} 
\usepackage{color}  
\usepackage{subfigure,bbm} 
\usepackage{amsmath,amsfonts,bm}
\usepackage{xspace}  
\usepackage{algorithm}

{\begin{list}{$\bullet$}{%
    \setlength{\topsep}{0in}
    \setlength{\partopsep}{0in}
    \setlength{\itemsep}{0in}
    \setlength{\parsep}{0in}
    \setlength{\leftmargin}{3.5em}
    \setlength{\rightmargin}{0in}
    \setlength{\itemindent}{0in}
}
}%
{\end{list}}

\def\eqref#1{equation~\ref{#1}}

\def\1{\bm{1}}

\DeclareMathAlphabet{\mathsfit}{\encodingdefault}{\sfdefault}{m}{sl}
\SetMathAlphabet{\mathsfit}{bold}{\encodingdefault}{\sfdefault}{bx}{n}

\def\mbw{\mathbf{w}}

\def\mbB{\mathbf{B}}

\def\mbE{\mathbf{E}}
\def\mbF{\mathbf{F}}

\def\mbP{\mathbf{P}} 
\def\mbQ{\mathbf{Q}}

\def\mbT{\mathbf{T}}

\def\mbW{\mathbf{W}}

\def\mbT{\boldsymbol{\mathcal{T}}}

\def\bxi{\boldsymbol{\xi}}

\def\bmu{\boldsymbol{\mu}}

\def\bchi{\boldsymbol{\chi}}

\def\bSigma{\boldsymbol{\Sigma}} 
 
\def\bTheta{\boldsymbol{\Theta}}

\def\mbw{\mathbf{w}}

\def\mbP{\mathbf{P}}
 
\def\mbB{\mathbf{B}}

\def\mbE{\mathbf{E}}
\def\mbF{\mathbf{F}}

\def\mbQ{\mathbf{Q}} 

\def\mbT{\mathbf{T}}

\def\mbW{\mathbf{W}}

\def\mbT{\boldsymbol{\mathcal{T}}}

\def\md{\mathcal{D}}

\def\mn{\mathcal{N}}

\def\mp{\mathcal{P}}   
   
\def\mcJ{\mathcal{J}}

\def\mbT{\mathbf{T}}

\def\bx{\boldsymbol{x}}  
  
\def\bs{\boldsymbol{s}}

\def\bt{\boldsymbol{t}}

\def\bc{\boldsymbol{c}}  
\def\bm{\boldsymbol{m}}

\def\bK{\boldsymbol{K}} 

\def\bX{\boldsymbol{X}} 
\def\bY{\boldsymbol{Y}}

\def\bD{\boldsymbol{D}}
\def\bE{\boldsymbol{E}}
\def\bF{\boldsymbol{F}}

\def\bK{\boldsymbol{K}}

\def\bM{\boldsymbol{M}}

\def\bxi{\boldsymbol{\xi}}

\def\bmu{\boldsymbol{\mu}} 
\def\bSigma{\boldsymbol{\Sigma}} 
 
\def\bTheta{\boldsymbol{\Theta}}

\def\bchi{\boldsymbol{\chi}}

\newcommand{\ea}[1]{\begin{equation}\begin{aligned}\centering \small #1 \end{aligned}\end{equation}}

\begin{document}   

\title{
\LARGE\bf
Learning a Shape-adaptive Assist-as-needed Rehabilitation Policy from Therapist-informed Input 
}  

\author{
Zhimin~Hou$^{\dag}$, 
Jiacheng~Hou$^{\dag}$, 
Xiao~Chen, 
Hamid~Sadeghian, 
Tianyu~Ren$^{*}$, 
Sami Haddadin
\thanks{
The authors would like to thank the Federal Ministry of Research, Technology, and Space (BMFTR) for its support as part of the research program Communication Systems "Souverän. Digital. Vernetzt.". 
Joint project 6G-life, project identification number: 16KIS2414.
}
\thanks{$\dag$ The authors contribute equally.}  
\thanks{Zhimin~Hou is with Division of Industrial Data Science~(DIDS), School of Data Science, Lingnan University, Hong Kong, and Mechanical and Aerospace Engineering, Raleigh, NC 27695, USA.} 
\thanks{Jiacheng Hou is with School of Mechanical Engineering, Tongji University.} 
\thanks{Xiao Chen, Hamid Sadeghian, Tianyu Ren, Sami Haddadin are with the Chair of Robotics and Systems Intelligence, MIRMI-Munich Institute of Robotics and Machine Intelligence, Technical University of Munich, Germany.} 
\thanks{Sami Haddadin is also with Mohamed bin Zayed University of Artificial Intelligence, Abu Dhabi, UAE.} 
\thanks{$*$ Corresponding author: Tianyu Ren (e-mail: tianyu@robot-learning.de).} 
} 

\maketitle  
\begin{abstract} 
Therapist-in-the-loop robotic rehabilitation has shown great promise in enhancing rehabilitation outcomes by integrating the strengths of therapists and robotic systems. 
However, its broader adoption remains limited due to insufficient interaction and limited adaptation capability. 
This article proposes a novel telerobotics-mediated framework that enables therapists to intuitively deliver assist-as-needed~(AAN) therapy based on two primary contributions. 
First, our framework encodes the therapist-informed corrective force into via-points in a latent space, allowing therapists to provide only minimal assistance while encouraging patients maintaining their own motion preferences. 
Second, a shape-adaptive AAN rehabilitation policy is learned to partially and progressively deform the reference trajectory for movement therapy based on encoded patient motion preferences and therapist-informed via-points. 
The effectiveness of the proposed shape-adaptive AAN strategy was validated on a telerobotic rehabilitation system using two representative tasks. 
The results demonstrate its practicality for remote AAN therapy and its superiority over two state-of-the-art methods in reducing corrective force and improving movement smoothness. 
\end{abstract}  

\section{Introduction}\label{sec-intro} 
Rehabilitation robots can provide precise and intensive therapies for patients with neurological injuries, offering a promising solution to the shortage of healthcare resources~\cite{hasson2023neurorehabilitation}. 
Numerous robotic systems~\cite{perry2007upper,hou2025bilateral} and control methodologies~\cite{proietti2016upper},\cite{yang2022task} have been developed to promote robot-assisted rehabilitation. 
Particularly, adaptive control and machine learning techniques have shown promising results in programming flexible therapy strategies tailored to patients' individual needs, known as \textit{assist-as-needed}~(AAN) strategies~\cite{pehlivan2015minimal,pezeshki2023cooperative}. 
Nevertheless, the flexibility of existing AAN strategies cannot match the capability of a skilled therapist due to the difficulty of acquiring sufficient knowledge and programming experience-based  adjustments~\cite{shahbazi2016robotics},\cite{liu2022home}. 
To address this limitation, \textit{therapist-in-the-loop} robotic rehabilitation has emerged as a paradigm that integrates therapists' expertise into robot controllers~\cite{atashzar2018computational},\cite{luciani2024therapists}. 
However, most existing robotic systems still struggle to perform effective therapist-in-the-loop rehabilitation training because of their limited interaction and adaptation capabilities. 

First, current robotic rehabilitation systems lack effective mutual interaction to safely integrate therapists' expertise into robot execution~\cite{wolbrecht2008optimizing},\cite{asl2021satisfying}. 
Robot controllers are typically designed to track the predefined reference trajectories based on the measured interaction forces and actual trajectories for desired rehabilitation goals~\cite{shahbazi2016robotics}. 
Therapists, in turn, contribute by setting rehabilitation goals and modulating the therapy strategies tailored to individual patients for robot execution based on previous expertise in evaluating patient status~\cite{liu2022home}. 
Therapists typically demonstrate therapeutic movements/forces for robot-mediated rehabilitation using the same robot in a teach-and-play mode~\cite{najafi2020using}. 
By contrast, telerobotics-mediated systems enable therapist and patient to operate separate devices, enabling real-time mutual interaction~\cite{chen2023tele},\cite{luciani2024therapists}. 
Therapists can utilize a haptic device to perceive the interaction force between the patient and the robot~\cite{shahbazi2016robotics}. Additional sensors and estimation methods were employed to achieve concurrent demonstration and therapy~\cite{liu2020home}. 
However, no current telerobotic-mediated systems can intuitively incorporate therapist's input into robot control without relying on reprogramming. 

\begin{figure*}[!t]  
\setlength{\abovecaptionskip}{-0.05cm}  
\centering  
{\includegraphics[width=1.0\linewidth]{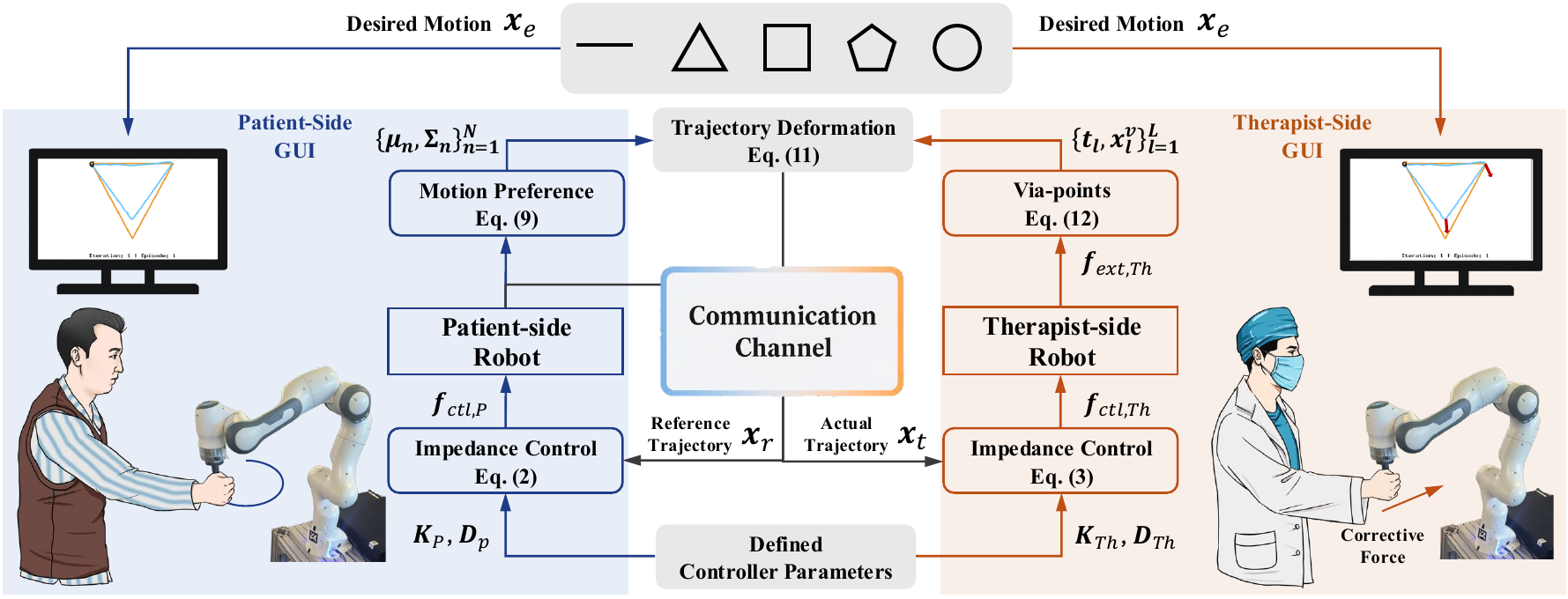}}   
\caption{
Overview of learning our shape-adaptive AAN rehabilitation policy based on two isomorphic robots interacting with patient and therapist through the end-effector, separately. 
The therapist first selects a specific rehabilitation goal, whose desired motion is displayed on both patient-side and therapist-side GUIs. 
Two impedance controllers with pretested control parameters are implemented for patient-side and therapist-side robots to repetitively complete the therapy. 
On the patient side, the reference trajectory is iteratively deformed during movement therapy. The patient motion preferences is encoded from previously collected actual trajectories. 
On the therapist side, another impedance controller is implemented to reproduce the actual motion of patient in real time, allowing the therapist to apply the corrective forces. 
The via-points are extracted for reference trajectory deformation, which are the transmitted to patient-side robot for the implementation of next therapy iteration. 
}
\label{fig:overview-framwork} 
\vspace{-0.8cm}  
\end{figure*}   
Second, current robotic rehabilitation strategies struggle to deliver effective AAN therapy due to their limited adaptability. 
The key idea of developing AAN training strategies is to encourage active participation by only providing the necessary level of assistance~\cite{shahbazi2016robotics,dalla2021review,yang2022task}. 
Patients' voluntary output forces and motion tracking errors are commonly used as metrics to assess patient engagement and guide the design of robot controllers for various rehabilitation goals~\cite{warraich2010neural,pehlivan2015minimal,asl2020field}. 
While therapists' expertise allows for comprehensive evaluation of the patients' status through multi-modal feedback, therapists' corrective actions are essential to promote therapist-level adaptive assistance~\cite{sharifi2020assist}. 
Numerous studies have leveraged probabilistic models to reproduce therapist expertise for shaping reference trajectories and force profiles in AAN therapy~\cite{proietti2016upper}. 
In these methods, rehabilitation goals are defined by the therapist and the corresponding references are generated for robot controllers~\cite{pezeshki2023cooperative,yang2022task}. 
However, the effectiveness of therapy remains constrained by the difficulty of balancing robotic assistance with patients' active participation. Therefore, a promising therapist-in-the-loop AAN training strategy that maximizes patient engagement while seamlessly integrating therapist-informed corrections remains necessary. 

This article proposes a novel therapist-in-the-loop rehabilitation framework that leverages two isomorphic collaborative robots to deliver AAN therapy. 
The contributions are twofold. First, therapist-informed corrective forces are encoded as via-points in a latent space, enabling the therapist to provide only the minimal necessary assistance instead of re-planning the entire reference trajectory or force profile. 
Second, a shape-adaptive AAN rehabilitation policy is learned to partially and progressively deform the reference trajectory based on the encoded patient motion preferences and therapist-informed via-points. 

\section{Related Works} 
\subsection{AAN Training Strategies}\label{subsec-aan-strategies}  
AAN training strategies have been investigated at multiple levels for patients with various rehabilitation goals by modulating the amount of assistance according to the biomechanical status of the patient or task performance~\cite{warraich2010neural},\cite{agarwal2017subject}. 
At the force level, adaptive controllers have been developed to successfully modify the level of assistance in real time~\cite{pehlivan2015minimal},\cite{pehlivan2014subject}. 
Furthermore, multi-modal adaptive controllers have been proposed to provide flexible assistive force across multiple training modes~\cite{li2020performance}. 
Beyond adjusting the assistive force, impedance control, a compliant control method, was utilized to regulate the patient-robot interaction dynamics during movement therapy~\cite{pezeshki2023cooperative}. 
Variable impedance controllers have been commonly developed to enable impedance-level AAN therapy, where individual reference stiffness parameters are predefined according to patient's task performance~\cite{pezeshki2023cooperative},\cite{huo2022adaptive},\cite{pareek2023ar3n}. 
These methods mainly achieved the force-level and impedance-level AAN therapy given a fixed reference trajectory.  
By contrast, the individualized reference trajectory deformation for motion-level AAN therapy has been proven to improve the stability~\cite{losey2017trajectory}. 
The reference trajectory was modified based on optimization-based deformation for patients with different mobility disabilities and at different recovery stages~\cite{Xu2023dmpmirror}.  
Additionally, probabilistic learning methods, Gaussian Mixture Model~(GMM) and Gaussian Mixture Regression~(GMR)~\cite{atashzar2018computational}, were applied to reproduce the therapist-demonstrated reference trajectories~\cite{shahbazi2016robotics}. 
More recently, moving beyond single rehabilitation tasks, task-level AAN strategies have been explored. 
Linear Gaussian policies~\cite{hou2025contextual} or neural network–based policies~\cite{zou2020learning} were developed to encode mappings from task contexts to a latent space of lower-level robot policies, which are formulated based on Kernelized Movement Primitive~(KMP)~\cite{huang2019kernelized}. 
Unfortunately, most existing approaches cannot match the flexibility of skilled therapists. 
An ideal AAN training strategy should combine the strengths of therapists’ expertise with the precise execution of robots. Crucially, the therapist should provide only the minimal necessary guidance, thereby maximizing patients’ active participation. 
\subsection{Therapist-in-the-loop Rehabilitation Training} 
Therapist-in-the-loop training strategies rely on demonstrating or modifying force profiles, therapeutic movements and therapeutic intensity following a teach-and-play paradigm~\cite{shahbazi2016robotics,luciani2024therapists}. 
During the demonstration phase, the therapeutic movements or force profiles delivered by therapists are recorded and fitted by probabilistic models~\cite{atashzar2018computational},\cite{luciani2024therapists}. 
In the reproduction phase, the learned probabilistic model can generate reference trajectories or force profiles without the therapist in the loop~\cite{najafi2020using}. 
The therapeutic intensity is modulated by modifying controller parameters of the patient-side robot to perform AAN therapy~\cite{yang2022task}. 
Additionally, leader-follower robotic systems have been developed for telerobotic-mediated rehabilitation to enable  concurrent demonstration and therapy. 
Therapists use a haptic device to demonstrate reference trajectories and to perceive patient-side interaction forces~\cite{liu2022home}, allowing for adjusting therapeutic movements and force profiles in real-time to ensure effective therapy~\cite{shahbazi2016robotics}. 
EMG sensors were employed to infer controller parameters for modulating therapy intensity. 
The telerobotics-mediated rehabilitation system enables the decoupling of evaluation and therapeutic treatment. 
Unfortunately, in existing strategies, therapists are typically required to modulate entire therapeutic movements or force profiles~\cite{luciani2024therapists}. 
By contrast, an ideal AAN therapy strategy would encode therapist-informed corrective forces to guide the movement therapy.  
\section{Overview Framework}\label{sec-problem}  
The objective of the proposed framework is to enable therapists to provide the necessary guidance for delivering the AAN therapy to patients. 
As illustrated in Fig.~\ref{fig:overview-framwork}, the telerobotic system consists of two identical collaborative robots to perform the therapist-in-the-loop training. 
One robot, located at the patient side~(patient-side robot), is driven by an impedance controller to guide the patient during movement therapy. 
The other robot, located on the therapist side~(therapist-side robot), is driven by another impedance controller, allowing him/her to perceive the actual trajectories of the patient-side robot. Two graphical user interfaces~(GUIs) are developed to enable mutual visual interaction between therapists and patients. For AAN therapy, the reference trajectory is adapted to encourage active participation of patients where the patient motion preferences are maintained. 
Furthermore, the therapist-side robot isolates the therapist-informed corrective force from the patient-side interaction force, enabling it to provide accurate guidance. 

This article focuses on movement therapy given a set of predefined desired motions~\cite{pareek2023ar3n}. 
Once a rehabilitation task is selected by the therapist~(see Fig.~\ref{fig:overview-framwork}), the desired motion $\bx_e$ is displayed by GUIs to provide task instructions for both therapists and patients. 
The robot-assisted rehabilitation relies on repetitive therapy based on two lower-level interactive controllers depicted in Section~\ref{subsec-interactive-control}. 
During each therapy iteration, the patient's actual trajectories $\bx_t$ are measured from the patient-side robot. 
The patient motion preferences are then encoded from recent actual trajectories by learning a probabilistic model~(see Section~\ref{subsec-motion-preference-encoding}). 
To minimize therapist assistance and encourage patient active participation, therapist-informed via-points are extracted from the measured corrective force, rather than transmitting the entire force profile or reference trajectory for patient-side robot control. 
The corresponding reference trajectory $\bx_r$ for subsequent therapy iteration is then partially deformed according to the therapist-informed via-points, while preserving the patient's motion preferences in the remaining segments~(see Section~\ref{subsec-trajectory-deformation}). 
Finally, the therapist's therapeutic skills are encoded as via-points in a latent space. 
A partial least squares regression function is fitted to reproduce these therapist-informed therapy skills~(see Section~\ref{subsec-stiffness-regulation}). 

\section{Method}\label{sec-method}  
\vspace{-0.2cm}
\subsection{Lower-level Interactive Control}\label{subsec-interactive-control} 
The robot dynamics in $n$ Degree of Freedom~(DoF) Cartesian space are given by
\begin{equation}
\begin{aligned}
\bM_{C,i} (\bx_{i}) \Ddot{\bx}_{i} &+ \bc_{C,i}(\bx_{i}, \dot{\bx}_{i}) + \boldsymbol{g}_{C,i}(\bx_{i}) \\
&= \boldsymbol{f}_{cmp,i} + \boldsymbol{f}_{ctrl,i} + \boldsymbol{f}_{ext,i} 
\end{aligned}
\label{equ:robot_dynamics}
\end{equation}
where $\bM_{C,i} \in \mathbb{R}^{n \times n}$ denotes the inertia matrix of the robot, and $\bc_{C,i} \in \mathbb{R}^n$ and $\boldsymbol{g}_{C,i} \in \mathbb{R}^n$ denote the Coriolis/centrifugal and the gravity vector, respectively. $\boldsymbol{f}_{cmp,i} \in \mathbb{R}^n$ is the control force to compensate the Coriolis/centrifugal and the gravity vectors. $\boldsymbol{f}_{ctrl,i} \in \mathbb{R}^n$ is the control command. 
Moreover, $\boldsymbol{f}_{ext,i} \in \mathbb{R}^n$ denotes the external force applied by human user to the robot. The subscript $-_i$ is indexed by $i \in \{P, Th\}$, where $-_P$ indicates the robot at the patient side, and $-_{Th}$ for the therapist-side robot. 

\subsubsection{Patient-side Robot Control} 
On the patient side, the robot aims to track the reference trajectory generated by the therapist input~\cite{chen2023tele}, thus the controller at the patient side is designed as, 
\begin{equation}
\boldsymbol{f}_{ctrl,P} = \bK_P (\bx_{r} - \bx_P) + \bD_P (\dot{\bx}_{r} - \dot{\bx}_P), 
\end{equation}
where the $\bK_P, \bD_P \in \mathbb{R}^{n \times n}$ are  positive definite stiffness matrices and damping matrix, respectively. And the $\bx_{r}$ is the generated reference trajectory. 

\subsubsection{Therapist-side Robot Control} 
The robot on the therapist side aims to track the patient-side robot so that the therapist understand the situation of the patient~\cite{chen2023tele}. 
Thus the controller on the therapist-side robot is designed as,
\begin{equation}
\label{equ-patient-side-impedance-robot}
\boldsymbol{f}_{ctrl,Th} = \bK_{Th} (\bx_P - \bx_{Th}) + \bD_{Th} (\dot{\bx}_P - \dot{\bx}_{Th}), 
\end{equation}
where the $\bK_{Th}, \bD_{Th} \in \mathbb{R}^{n \times n}$ are the positive definite stiffness and damping matrices for the therapist-side robot, respectively. 
And the interaction force between the therapist $\boldsymbol{f}_{ext,Th}$ will be the input of the trajectory generation. 
\begin{figure}[!t]  
\setlength{\abovecaptionskip}{-0.05cm}  
\centering  
{\includegraphics[width=1.0\linewidth]{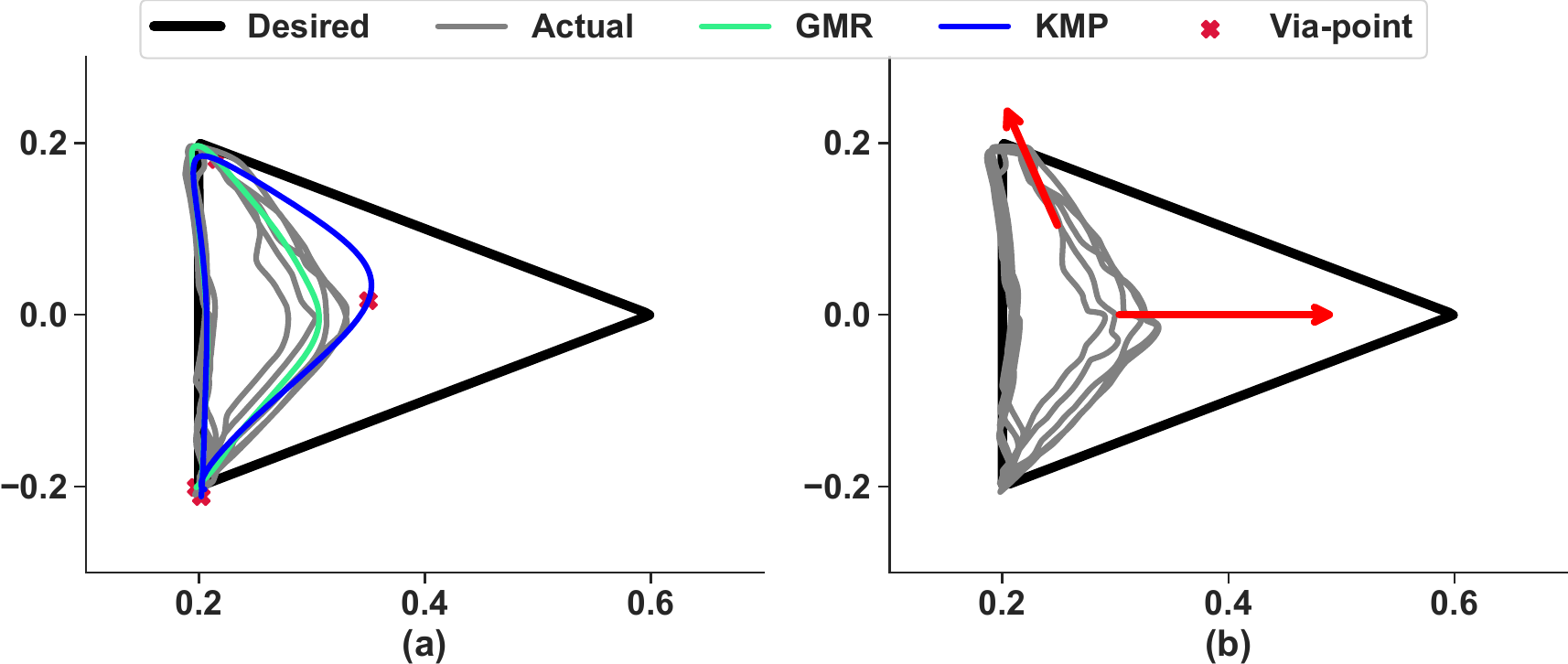}}   
\caption{
Illustration of patient motion preferences encoding, therapist-informed via-point extraction, and reference trajectory deformation in the proposed framework.~(a)~Gray lines are patient's actual motion trajectories. The green line indicates the estimated mean motion trajectory according to Section~\ref{subsec-motion-preference-encoding}. Red scatter points indicate therapist-informed via-points inferred from the therapist-informed corrective force and the desired motion. The blue line indicates the generated reference trajectory according to Section~\ref{subsec-trajectory-deformation}.~(b)~Gray lines represent actual motion trajectories reproduced by therapist-side robot, while therapist applies the effective corrective force as the red arrows to inform the via-points. 
}  
\label{Fig2:trajectory-deformation} 
\vspace{-0.6cm}  
\end{figure} 
\subsection{Patient Motion Preferences Encoding}\label{subsec-motion-preference-encoding} 
For each therapy iteration, the therapist-side impedance controller in (\ref{equ-patient-side-impedance-robot}) is used to perform the movement therapy by tracking a reference trajectory $\bx_r$ for $J$ times. 
Most recent actual trajectories of patient~(e.g. gray lines in the Fig.~\ref{Fig2:trajectory-deformation}(a)) are collected and down-sampled into a set of waypoints with length $N$ for learning a probabilistic model to encode patient motion preferences~\cite{2025iroskmp,hou2026teachingbot}. 
The dataset $\md_P$ is constructed as $\md_P = \{\{ \bxi_{n,j}^i, \bxi_{n,j}^o \}_{n=1}^N\}_{j=1}^{J}$. 
$\bxi_{n,j}^i \in \mathbb{R}^{d_i}$ and $\bxi_{n,j}^o \in \mathbb{R}^{d_o}$ represent the input and output vectors with dimensions of $d_i$ and $d_o$, respectively. 
This study focuses on time-driven trajectories for movement therapy, where the input $\bxi_{n,j}^i$ corresponds to time $\bt$, and the output $\bxi_{n,j}^o$ corresponds to the end-effector position $\bx$ of the patient-side robot. 
GMMs are used to encode the mean and variance of motion preferences using collected waypoints in $\md^i_P$. The covariance reflects the variability of the actual trajectories~\cite{zou2020learning}. 
The resulting joint probability distribution of time $\bt$ and position $\bx$ is given as $\mp(\bt, \bx) = \sum\nolimits_{c=1}^C \varpi_c \mn(\bt, \bx | \bmu_c, \bSigma_c)$. 
$C$ is the number of Gaussian components, which controls the smoothness of the fitted mean trajectory. 
The Gaussian component parameters $\{\bmu_c, \bSigma_c, \varpi_c\}_{c=1}^C$ are characterized by its mean $\bmu_c$, covariance $\bSigma_c$, and weight $\varpi_c$, estimated from $\md_P$ using the \textit{Expectation-Maximization} algorithm~(see green line and ellipses in Fig.~\ref{Fig2:trajectory-deformation}(a)). 
The conditional probabilistic trajectory, denoted by $\{\widehat{\bx}_n\}_{n=1}^N$, is then retrieved via GMR as $\mp(\widehat{\bx}_n | \bt_n) \sim \mn(\widehat{\bmu}_n, \widehat{\bSigma}_n)$. $\widehat{\bmu}_n$ and $\widehat{\bSigma}_n$ are the conditional mean and covariance that can be calculated from the estimated GMMs parameters. 

The parameterized potential trajectories of patient can be modeled as $\bx(\bt) = \bTheta(\bt)^T\mbw$~\cite{huang2019kernelized}. 
$\bTheta \in \mathbb{R}^{d_Bd_o \times d_o}$ is the basis feature matrix, $d_B$ is the dimensionality of the basis feature, and $\mbw \in \mathbb{R}^{d_Bd_o}$ is the weight vector, assumed to follow a normal distribution $\mbw \sim \mn(\bmu_{\mbw}, \bSigma_{\mbw})$ with unknown mean $\bmu_{\mbw}$ and covariance $\bSigma_{\mbw}$. 
The parameterized probability of trajectory distribution is then given by  
\ea{  
\mp_{\mbw}(\bx |\bt) = \mn(\bTheta(\bt)^T\bmu_{\mbw},\bTheta(\bt)^T\bSigma_{\mbw}\bTheta(\bt)) 
} 
where $\{\bmu_{\mbw},\bSigma_{\mbw}\}$ are the unknown mean and covariance to be estimated by minimizing the Kullback-Leibler~(KL) divergence objective as 
\ea{
\label{subsec-kmp-objective}  
\boldsymbol{\mcJ}(\bmu_{\mbw}, \bSigma_{\mbw} | \md_P) = \sum\nolimits_{n=1}^N D_{KL}(\mp_{\mbw}(\bx |\bt_n) || \mp_p(\bx | \bt_n)) 
}  
where $\mp_p(\bx | \bt_n)$ denotes the prior probability distribution of reference trajectory estimated from the dataset $\md_P$. 
Leveraging the properties of KL-divergence for Gaussian distributions, the mean and variance can be obtained by solving two sub-problems\cite{huang2019kernelized,zou2020learning}. 
\subsection{Therapist-informed Trajectory Deformation}\label{subsec-trajectory-deformation} 
The therapist-side robot can reproduce the actual trajectory of patient for $J$ times during each therapy iteration. 
The patient motion preferences are estimated from the previous dataset $\md_P^i$ as in Section~\ref{subsec-motion-preference-encoding}. 
For AAN therapy, the therapist applies corrective forces to partially deform the reference trajectory instead of updating the weight vector $\mbw$ to re-plan the entire reference trajectory. 
The therapist-informed corrective force $\boldsymbol{f}_{ext,Th}(\bt)$ is measured from the therapist-side robot during the reproduction of the patient's actual trajectory. 
When $\boldsymbol{f}_{ext,Th}(\bt)$ exceeds a pretested threshold $|\overline{\boldsymbol{f}}|$, it indicates that the therapist intends to correct the therapeutic movement of the patient-side robot. 
As shown in Fig.~\ref{Fig2:trajectory-deformation}(b), when a corrective force is activated at time $\bt_v$, a therapist-informed via-point is inserted at $\bt_v^{\prime} = \bt_v + \delta \bt$ to enable the partial deformation of reference trajectory. 
An additional dataset $\md_T$ that contains all therapist-informed via-points is introduced for each therapy episode to deform the reference trajectory for next therapy iteration. 
$\md_T$ shares the same structure as dataset $\md_P$ but consists of far fewer therapist-informed via-points~($L << N$) to encourage the active participation of patients at other time.  
When the therapist-informed via-points $\{\bt_l, \bx_l^v\}_{l=1}^L$ are assumed to follow a Gaussian distribution $\mp_v(\bx_l^v | \bt_l) \sim \mn(\widehat{\bmu}_l^v, \widehat{\bSigma}_l^v)$, the objective in (\ref{subsec-kmp-objective}) for estimating  $\{\bmu_{\mbw}, \bSigma_{\mbw}\}$ can be reformulated following the idea in \cite{huang2019kernelized}, as
\ea{ 
\label{equ-kmp-objective}
\boldsymbol{\mcJ}(\bmu_{\mbw}, \bSigma_{\mbw} | \md_P, \md_T) &= \sum\nolimits_{n=1}^N D_{KL}(\mp_{\mbw}(\bx |\bt_n) || \mp_p(\bx | \bt_n)) \\
 &+ \sum\nolimits_{l=1}^L D_{KL}(\mp_{\mbw}(\bx |\bt_l) || \mp_v(\bx | \bt_l))  
} 
where $\mp_p(\bx | \bt_n)$ and $\mp_v(\bx | \bt_l)$ denote the prior probability distribution estimated from datasets $\md_P$ and $\md_T$, respectively. 
To obtain a closed-form solution, as in (\ref{subsec-kmp-objective}), an extended dataset $\md_U = \md_P \cup \md_T$ is constructed, and the objective function is reformulated as 
\ea{
\label{subsec-kmp-final-objective} 
\boldsymbol{\mcJ}(\bmu_{\mbw}, \bSigma_{\mbw} | \md_U) = \sum\nolimits_{j=1}^{N+L} D_{KL}(\mp_{\mbw}(\bx | \bt_j) || \mp_u( \bx | \bt_j))  
} 
where the prior probability $\mp_u(\bx | \bt_j) \sim \mn(\bmu_j, \bSigma_j)$ is estimated from the extended dataset $\md_U$. 
Intuitively, the therapist-informed via-points are extracted from the corrective force and the desired motion $\bx_e$ of the selected rehabilitation task. 
The mean and covariance of each therapist-informed via-point at $\bt_v^{\prime}$ are designed as
\ea{
\label{equ-deriving-via-points}
\bmu(\bt_v^{\prime}) &= \beta_{\bmu} \cdot (\bx_e(\bt_v^{\prime}) - \bmu_{\mbw}(\bt_v^{\prime})) \cdot \frac{\boldsymbol{f}_{ext,Th}(\bt_v)}{|\boldsymbol{f}_{ext,Th}(\bt_v)|} + \bx_r(\bt_v^{\prime})\\ 
\bSigma(\bt_v^{\prime}) &= \bSigma_{\mbw}(\bt_v^{\prime})  
}  
where $\beta_{\bmu}$ is a scaling factor determining the amplitude of the reference trajectory deformation. 
Furthermore, to enable smooth complete trajectory deformation, two additional via-points at $\bt_s=0$ and $\bt_e=\Delta T$ are inserted at the start and end points. $\Delta T$ is the duration of each therapy episode. 
The mean values of these via-points and the corresponding covariance matrices are defined as $\{\bx_e(\bt_s), \bx_e(\bt_e)\}$ and $\{\bSigma(\bt_s), \bSigma(\bt_e) \}$. 
The blue line in Fig.~\ref{Fig2:trajectory-deformation}(b) represents the generated reference trajectory for the next therapy iteration, which preserves the patient’s previous motion preferences while adapting to the therapist-informed via-points. 
\vspace{-0.2cm} 
\subsection{Reproducing Therapist Input Skills}\label{subsec-stiffness-regulation} 
A latent space was constructed to represent the therapist's input $\boldsymbol{\mathcal{\chi}}_v$ with the dimension of $M$. At each $i-$th therapy iteration, the therapist state is denoted as $\bs_i$ with a dimension of $N$: 
\ea{
\label{equ-therapist-state}
\bs_i = \bmu_{\mbw}^i -\bchi_e, \in \mathbb{R}^N 
} 
where $\bmu_{\mbw}^i$ is the mean waypoints encoded from patient's actual trajectories and $\bchi_e$ is the extracted waypoints from desired motion $\bx_e$. 

A dataset was constructed as $\md_T = \{(\bs_i, \bx_{l,i}^v)_{i=1}^I\}$. 
A partial least squares~(PLS) regression function is utilized to reproduce the therapist input $\bY \in \mathbb{R}^{I \times M}$ from the collected feature $\bX \in \mathbb{R}^{I \times N}$, as, 
\ea{
\bY = \bX\mbB 
}
where $\mbB \in \mathbb{R}^{N \times M}$ is the weight matrix. $\bX$ and $\bY$ can be decomposed as, 
\ea{
\bX = \mbT\mbP^T + \bE, \bY = \mathbf{U}\mathbf{Q}^T + \bF
}
where $\mbP$ and $\mbQ$ are the loading matrices, $\mbE \in \mathbb{R}^{I \times N}$ and $\mbF \in \mathbb{R}^{I \times M}$ denote the residual matrices of $\bX$and $\bY$, respectively. When only using several latent dimensions, the weight matrix is derived as 
\ea{
\mbB = \mbW(\mbP^T\mbW)^{-1}\mbQ^T
}
where $\mbW$ is the weight matrix of the input matrix. 

\begin{algorithm}[t] 
\caption{Shape-adaptive AAN Rehabilitation Policy}   
\label{alg-aan-training}  
\begin{algorithmic}[1] 
\STATE Select rehabilitation task and desired motion $\bx_e$
\STATE Set hyperparameters $N$, $M$, $L$, $C$ 
\STATE Initialize controller parameters in Section~\ref{subsec-interactive-control} 
\STATE Initialize datasets $\md_P^0 \leftarrow \varnothing,~\md_T \leftarrow \varnothing$ 
\STATE Initialize weight matrix of PLS regression function $\mbB^0$  
\FOR {each therapy iteration $i=0$ \TO $I$}  
\STATE Estimate motion preference using $\md_P^i$(Section~\ref{subsec-motion-preference-encoding}) 
\STATE Collect therapist state $\bs_i$ according to (\ref{equ-therapist-state}) 
\STATE Collect therapist applied force $\boldsymbol{f}_{ext,Th}(\bt)$
\STATE Infer therapist-informed via-points $\bx_v^i$ using (\ref{equ-deriving-via-points}) 
\STATE Generate reference trajectory $\bx_r^i$ in Section~\ref{subsec-trajectory-deformation} 
\FOR {each therapy episode $j=0$ \TO $J$}    
\STATE Perform therapy following Section~\ref{subsec-interactive-control} 
\STATE Collect $\bx_t$ and store in $\md_P^i \leftarrow \md_P^i + \{\bx_t\}_{t=0}^T$ 
\ENDFOR 
\STATE $\md_T \leftarrow \md_T + \{(\bs_i, \bx_v^i)\}$, $\md_P^i \leftarrow \varnothing$ 
\ENDFOR  
\STATE Output optimized weight matrix: $\mbB^{\star}$  
\end{algorithmic} 
\end{algorithm} 
\section{Experimental Results}\label{sec-exp-results} 
\subsection{Experimental Testbed}\label{subsec-exp-setups} 
A telerehabilitation robotic system was set up as a testbed to validate the effectiveness of the proposed framework for upper-limb rehabilitation. 
The system consists of two identical 7-DoF Franka Emika Panda robots, controlled by two PCs running a real-time Linux kernel at the control frequency of 1000 Hz. 
Both the patient-side GUI and therapist-side GUIs were developed using PyQt5 to visualize the desired and actual motions as illustrated in Fig.~\ref{fig:overview-framwork}. 
Data transmission between the therapist-side PC and the patient-side PC was achieved using ZeroMQ within the same network, resulting in negligible communication delay. 
The hyperparameters of the lower-level interactive controllers used in Algorithm~\ref{alg-aan-training} are predefined and summarized in Table~\ref{tab-para-system-performance}.  
Without loss of generality, the robot in this work is constrained to Cartesian space; specifically, all rehabilitation tasks are executed within the $X–Y$ plane relative to the robot’s base frame~(see Fig.~\ref{fig:experimental-setups}). 
\vspace{-0.6cm} 
\subsection{Experimental Protocol}\label{subsec-experimental-protocol} 
\begin{figure}[!t]  
\setlength{\abovecaptionskip}{-0.00cm} 
\centering  
{\includegraphics[width=1.0\linewidth]{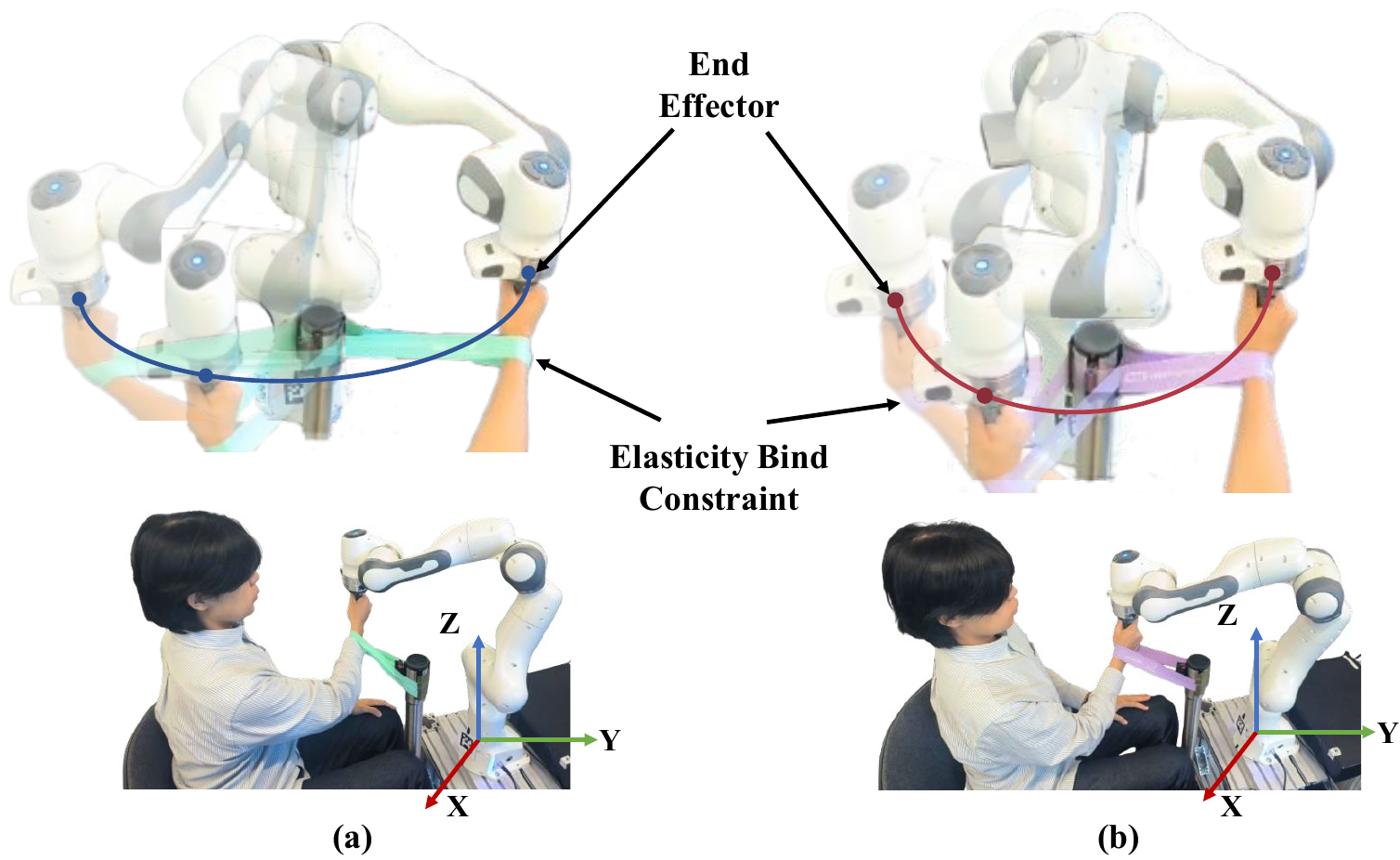}}    
\caption{
Illustration of the artificial motor disability induced using an elastic band to provide resistance.~(a)~Patient at Stage \#1 using a band with lower elasticity.~(b)~Patient at Stage \#2 using a band with higher elasticity. 
}
\label{fig:experimental-setups} 
\vspace{-0.2cm}    
\end{figure}

The objective of the subsequent human study is to validate the effectiveness of the proposed controller in incorporating the therapists' input for AAN therapy, instead of evaluating the biomechanical responses of the patient. 
Two able-bodied participants from the authors' laboratory were recruited to play the roles of therapist and patient. 
As illustrated in Fig.~\ref{fig:experimental-setups}, participants acting as patients were unable to follow the desired motions due to the impediment introduced by the elastic band. Moreover, different levels of band elasticity were applied to simulate various stages of patient recovery. 
Two rehabilitation tasks, each involving distinct motion patterns~\cite{shahbazi2016robotics}\cite{pareek2023ar3n}, were designed to demonstrate the effectiveness of the proposed controller in enabling AAN therapy. 
For each task, one participant acts as the therapist and is assumed to be able to accurately follow the desired motion, while another participant acts as the patient and receives AAN therapy. 
Accordingly, given the desired motion and predefined parameters, the therapist-informed AAN therapy was performed following the implementation in Algorithm~\ref{alg-aan-training}. 
Two baseline methods were also implemented to demonstrate the strength of our method. 
First, Baseline \#1 was implemented similar to \cite{losey2017trajectory},\cite{yang2022task}, the desired motion was selected as the reference trajectory and a variable impedance controller was applied to perform the movement therapy. 
Second, Baseline \#2 was implemented similar to \cite{luciani2024therapists}, the therapist-informed corrective force was directly transmitted to the patient-side robot whenever the therapist deems assistance necessary. 
Furthermore, two metrics were calculated for a fair comparison. 
First, the corrective force measured from patient-side robot was utilized to evaluate the robot's input and patient's active participation, denoted as $M_1$. 
Second, movement smoothness, a commonly used indicator of kinematic control, was calculated by spectral arc length~\cite{losey2017trajectory} to assess the quality of movement therapy, denoted as $M_2$. 
\begin{table}[!t]     
\centering    
\caption{Controller Parameters}   
\begin{tabular}{l | l}   
\hline   
$\bK_P = \textrm{diag}(200,200,1000, 50, 50, 50)$ & $C_1 = 10, C_2=10$ \\ 
$\bD_P = \textrm{diag}(10,10,57,13,13,13)$ & $|\overline{\boldsymbol{f}}|$=10N, $N$=200 \\
$\bK_{Th} = \textrm{diag}(800,800,800,50,50,50)$ & $L_1 = 4, L_2=5$ \\ 
$\bD_{Th} = \textrm{diag}(51,51,51,13,13,13)$ & $\lambda_{\mu}=1,\lambda_{\Sigma}=60,\varrho=2$ \\ 
$I=10, J=5$ & $\delta \bt =0.05s$, $\beta_{\bmu} = 1.0$ \\ 
\hline  
\end{tabular}  
\label{tab-para-system-performance}  
\vspace{-0.6cm}      
\end{table}   

\begin{table*}[t]
\centering
\setlength{\abovecaptionskip}{-0.02cm}   
\begin{tabular}{cccc}\hspace{-10pt}
\includegraphics[width=0.225\linewidth]{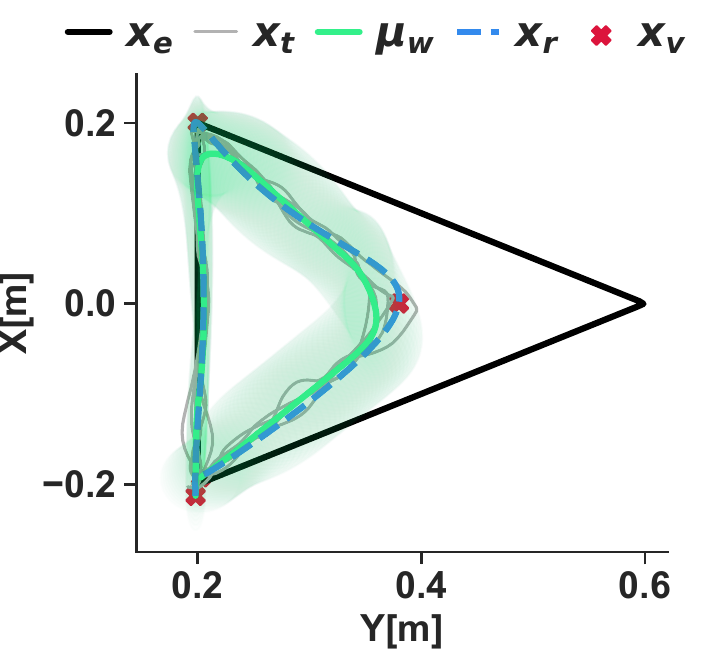}
\label{fig:sec3-inllustration-framework} & 
\includegraphics[width=0.225\linewidth]{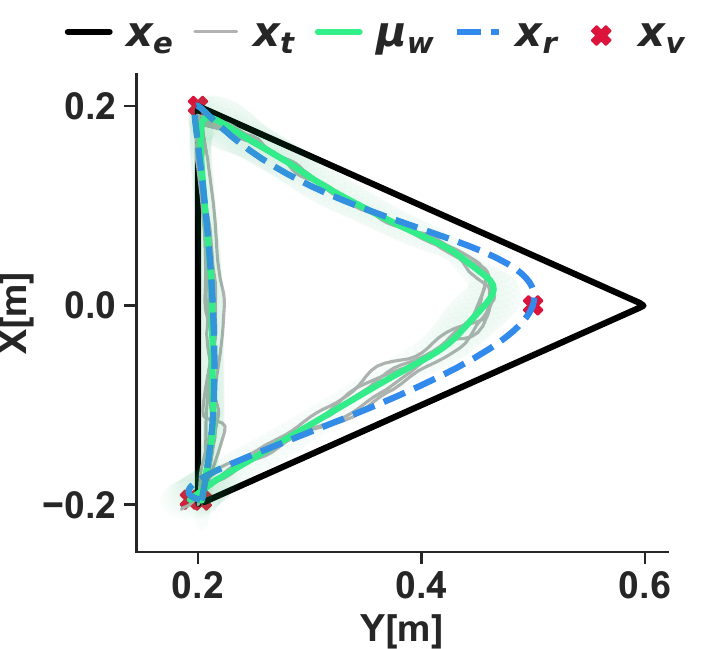}   
\label{fig:sec2-block-diagram} & 
\includegraphics[width=0.225\linewidth]{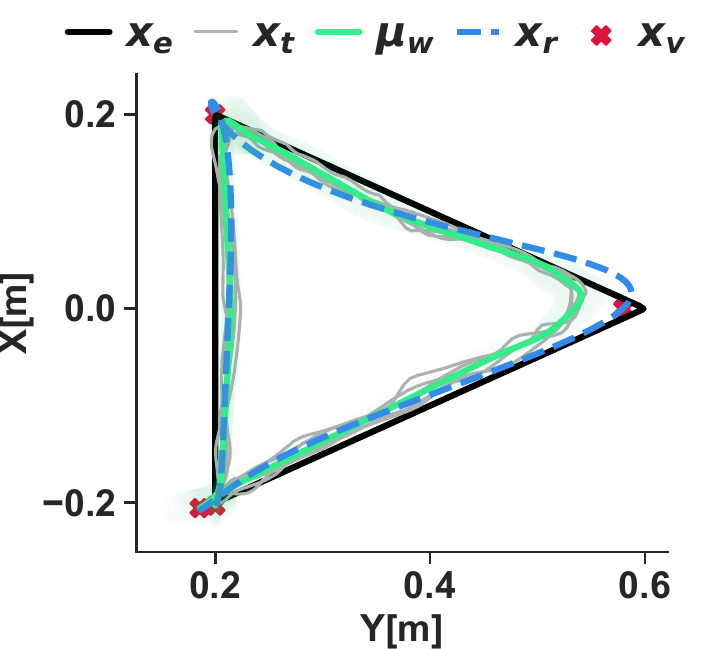}  
\label{fig:sec2-block-diagram} &
\includegraphics[width=0.225\linewidth]{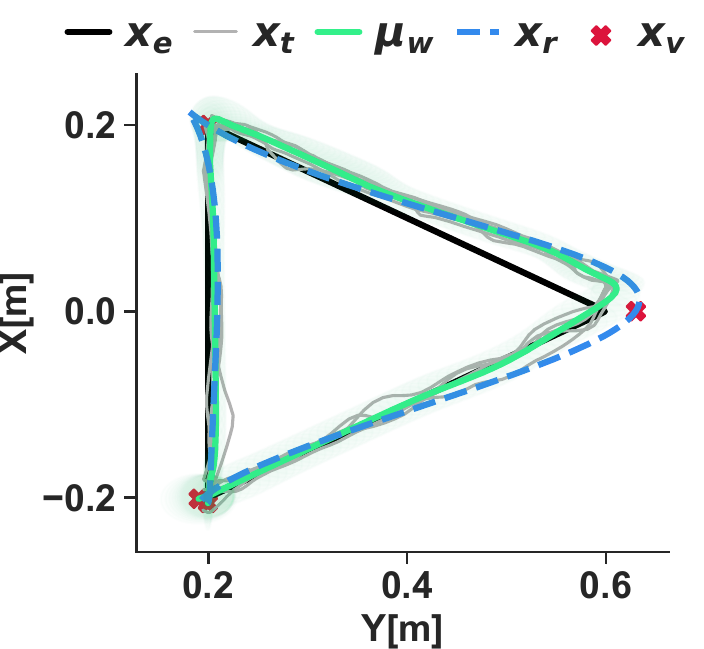}  
\label{fig:sec2-block-diagram} \\
\includegraphics[width=0.225\linewidth]{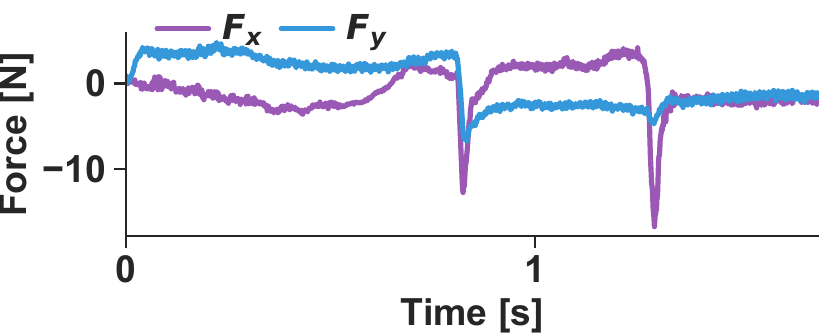}
\label{fig:sec3-inllustration-framework} & 
\includegraphics[width=0.225\linewidth]{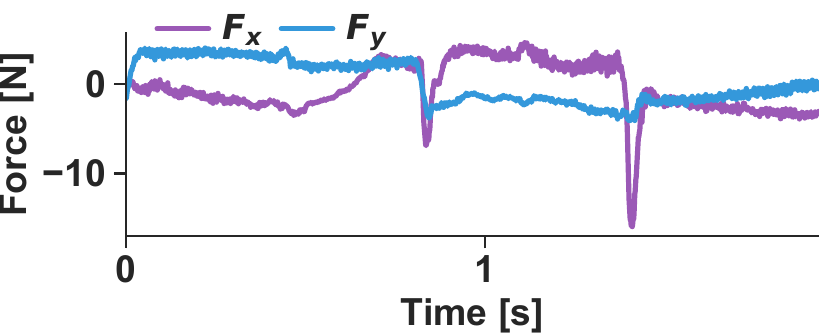}   
\label{fig:sec2-block-diagram} & 
\includegraphics[width=0.225\linewidth]{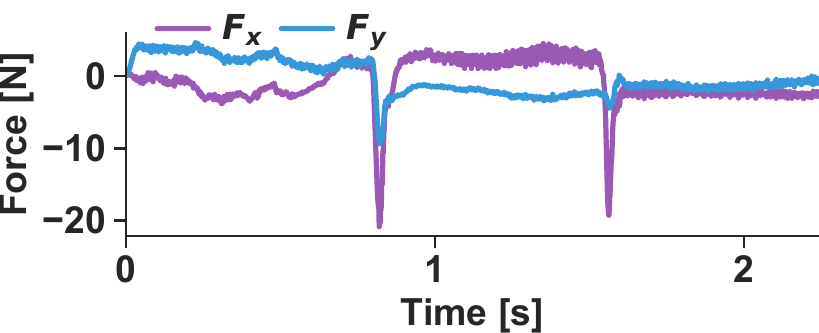}  
\label{fig:sec2-block-diagram} &
\includegraphics[width=0.225\linewidth]{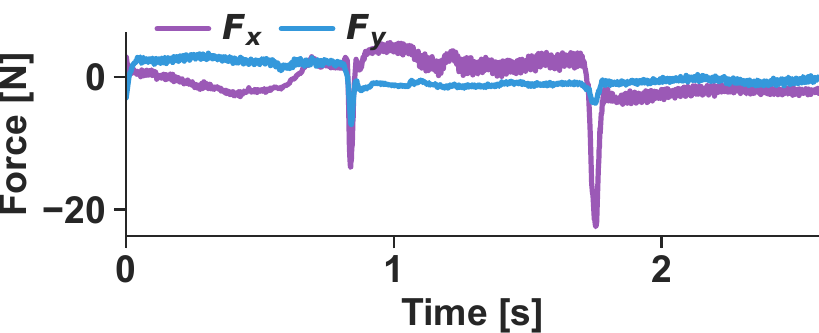}  
\label{fig:sec2-block-diagram} \\
(a) & (b) & (c) & (d) 
\end{tabular}  
\captionof{figure}{
Performance of Task \#1 by the patient at Stage \#1, where $x_e(t)$ denotes the desired motion, 
$x_r(t)$ represents the reference trajectory, 
$x_t(t)$ denotes the actual trajectory, 
$\mu_w(t)$ is the unknown mean of the learned motion primitive, 
and $\chi_v$ represents the therapist-informed via-point input. The first row illustrates the encoded patient motion preferences and the generated reference trajectories, while the second row shows the corrective forces applied by the therapist.~(a) Iteration 1; (b) Iteration 3; (c) Iteration 6; (d) Iteration 9. 
} 
\label{Fig2:theraputic-movement-task-1}  
\vspace{-0.4cm}     
\end{table*}  

\textbf{\begin{table*}[t]
\centering
\setlength{\abovecaptionskip}{-0.02cm}   
\begin{tabular}{cccc}\hspace{-10pt}
\includegraphics[width=0.225\linewidth]{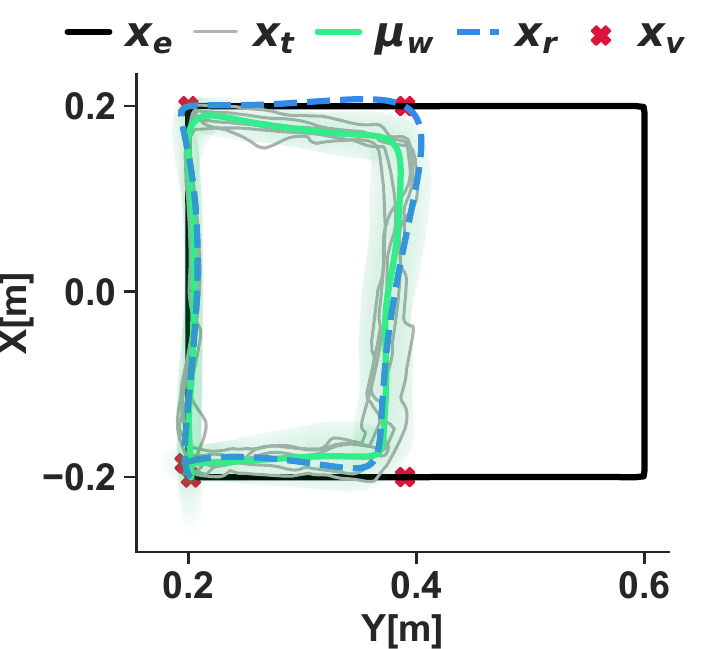}
\label{fig:sec3-inllustration-framework} & 
\includegraphics[width=0.225\linewidth]{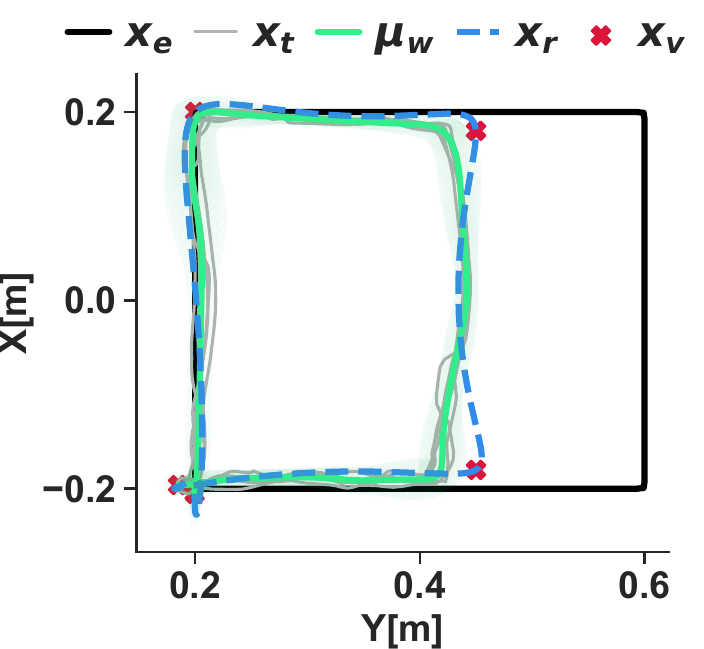}   
\label{fig:sec2-block-diagram} & 
\includegraphics[width=0.225\linewidth]{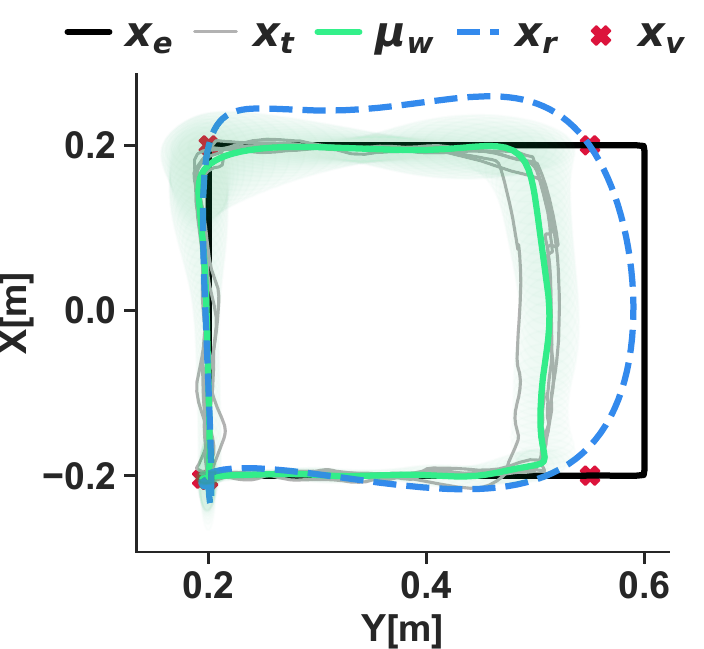}  
\label{fig:sec2-block-diagram} &
\includegraphics[width=0.225\linewidth]{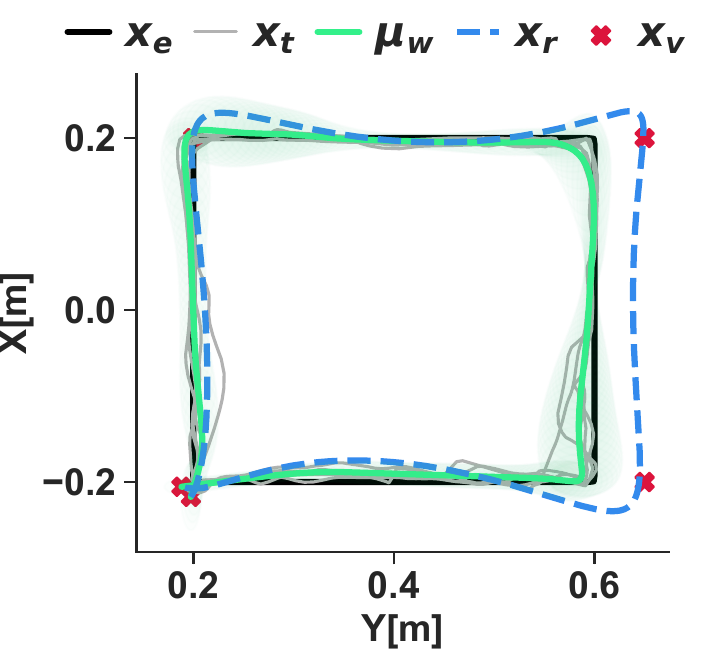}   
\label{fig:sec2-block-diagram} \\
\includegraphics[width=0.225\linewidth]{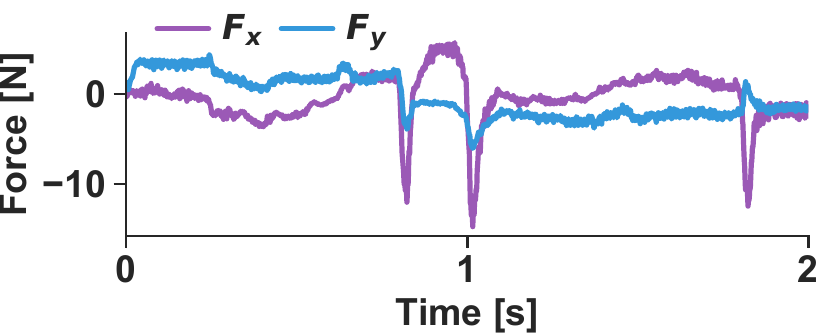}
\label{fig:sec3-inllustration-framework} & 
\includegraphics[width=0.225\linewidth]{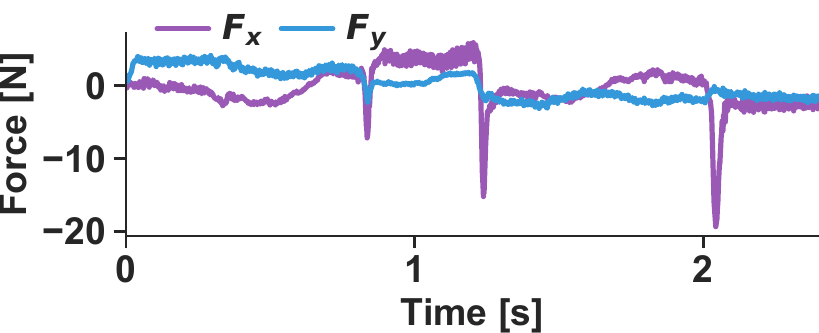}   
\label{fig:sec2-block-diagram} & 
\includegraphics[width=0.225\linewidth]{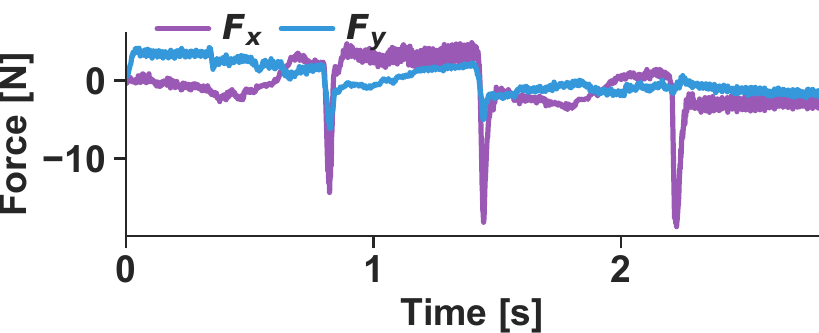}  
\label{fig:sec2-block-diagram} &
\includegraphics[width=0.225\linewidth]{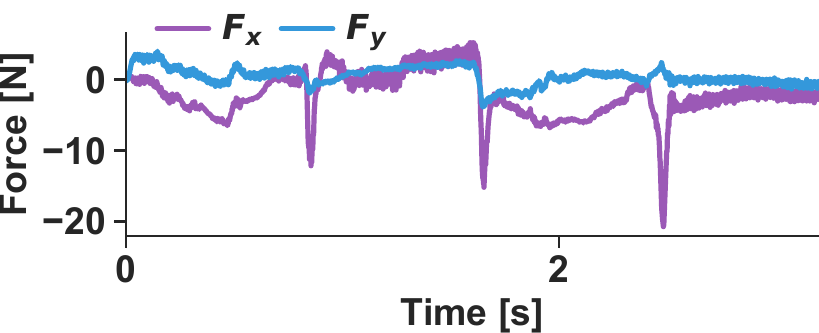}  
\label{fig:sec2-block-diagram} \\
(a) & (b) & (c) & (d) 
\end{tabular}  
\captionof{figure}{
Performance of Task \#2 by the patient at Stage \#1. The first row illustrates the encoded patient motion preferences and the generated reference trajectories, while the second row shows the corrective forces applied by the therapist.~(a) Iteration 1; (b) Iteration 3; (c) Iteration 6; (d) Iteration 9. 
} 
\label{Fig2:theraputic-movement-task-2}   
\vspace{-0.8cm}     
\end{table*}    
} 
\subsection{Experiment Results of Human Study}\label{subsec-human-study}  
The participant wearing an elastic band with low elasticity acted as the patient to perform two representative tasks~(see Fig.~\ref{fig:experimental-setups}(a)). 
Each task was repeated for $I=10$ therapy iterations without relying on any termination criteria. 
The learned patient motion preferences and generated reference trajectories are visualized in Fig.~\ref{Fig2:theraputic-movement-task-1} and Fig.~\ref{Fig2:theraputic-movement-task-2}. 
\subsubsection{Therapist-informed AAN Therapy} 
The desired motion of Task \#1~(a triangular trajectory) was visualized by black lines as shown in Fig.~\ref{Fig2:theraputic-movement-task-1}. 
At $i$-th therapy iteration, $J = 5$ actual therapy trajectories, visualized by gray lines, are collected to encode the motion preferences using GMMs with $C_1=10$. 
The green line and the shaded region represent the learned mean and variance. 
For a fair comparison, the therapist's skill is assumed to enable the patient to complete the key points of the desired motion. 
Therefore, $L_1=4$ via-points~(see Section~\ref{subsec-trajectory-deformation}) were extracted from the therapist-informed corrective force and visualized by red scatters. 
We can see that the therapist-informed corrective force plotted in the second row demonstrates that the therapist did not actively intervene in the current therapy session. 
Instead, the therapist only informed two via-points to guide the trajectory deformation for AAN therapy. 
The dotted blue line was the generated reference trajectory for next therapy iteration. 
\begin{figure*}[!t] 
\setlength{\abovecaptionskip}{-0.15cm}   
\centering  
\subfigure[]{\includegraphics[width=0.235\linewidth]{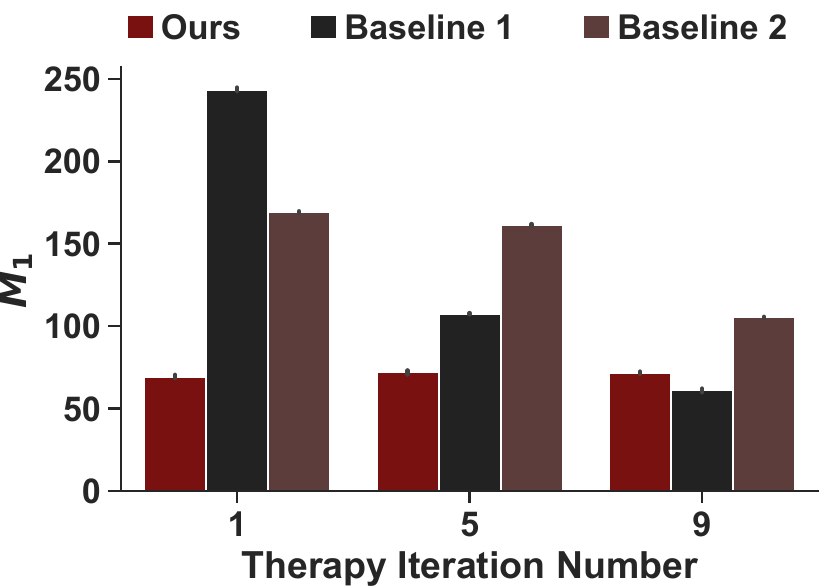}   
\label{Fig6:-metric1-task1}}   
\subfigure[]{\includegraphics[width=0.236\linewidth]{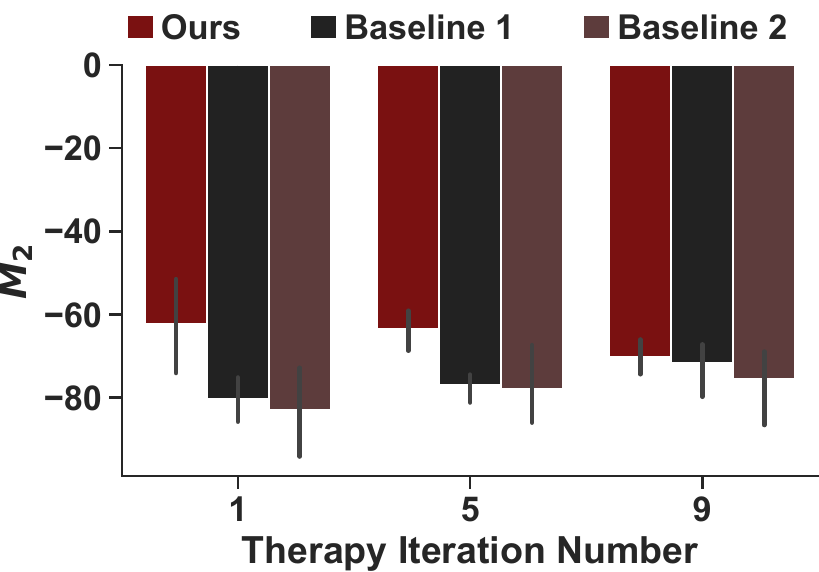}  
\label{Fig6:-metric2-task1}}  
\subfigure[]{\includegraphics[width=0.235\linewidth]{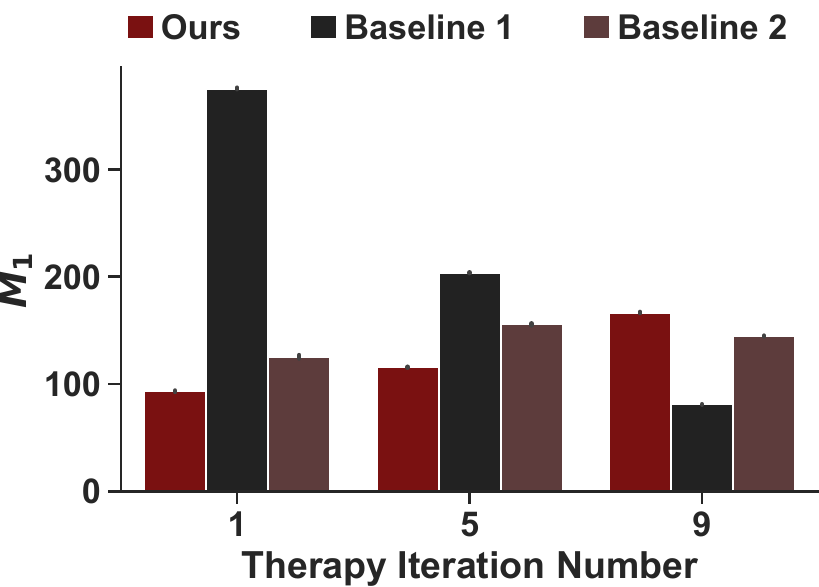}   
\label{Fig6:-metric1-task2}}   
\subfigure[]{\includegraphics[width=0.236\linewidth]{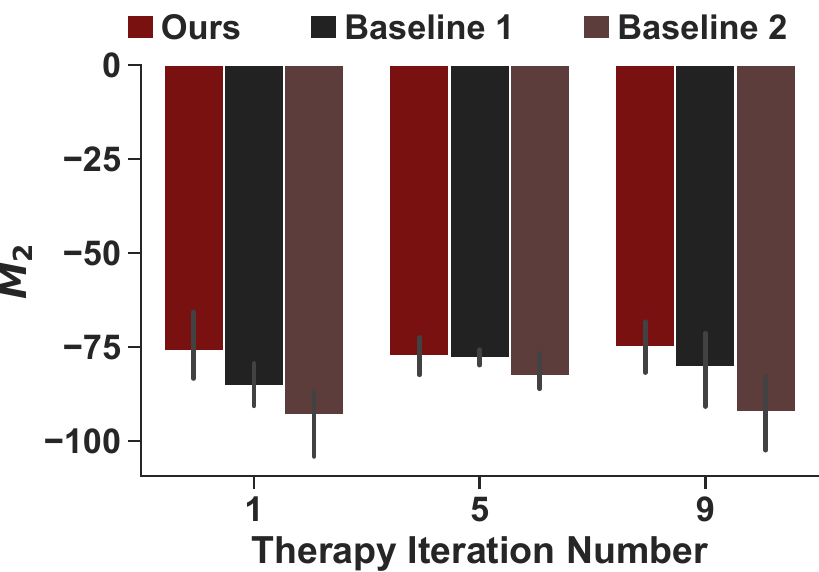}  
\label{Fig6:-metric2-task2}} 
\caption{
Comparison results of our controller and two baseline methods over three therapy iterations.~(a)~Average corrective force across five episodes for Task \#1;~(b)~Average movement smoothness index across five episodes for Task \#1;~(c)~Average corrective force across five episodes for Task \#2;~(d)~Average movement smoothness across five episodes for Task \#2. 
} 
\label{Fig6:comparison-with-baselines}   
\vspace{-0.6cm} 
\end{figure*} 
\begin{figure*}[!t] 
\setlength{\abovecaptionskip}{-0.15cm}   
\centering  
\subfigure[]{\includegraphics[width=0.23\linewidth]{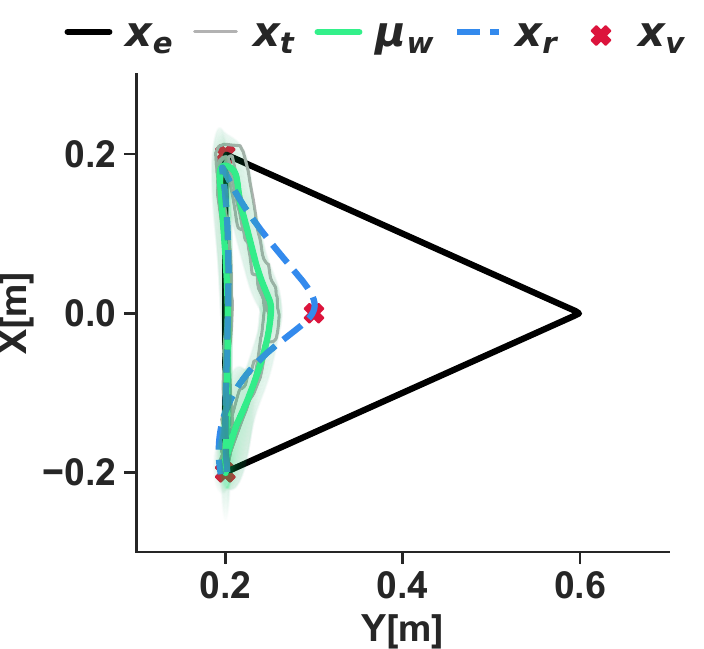}   
\label{Fig4(a):path-encoding-gmm}} 
\subfigure[]{\includegraphics[width=0.23\linewidth]{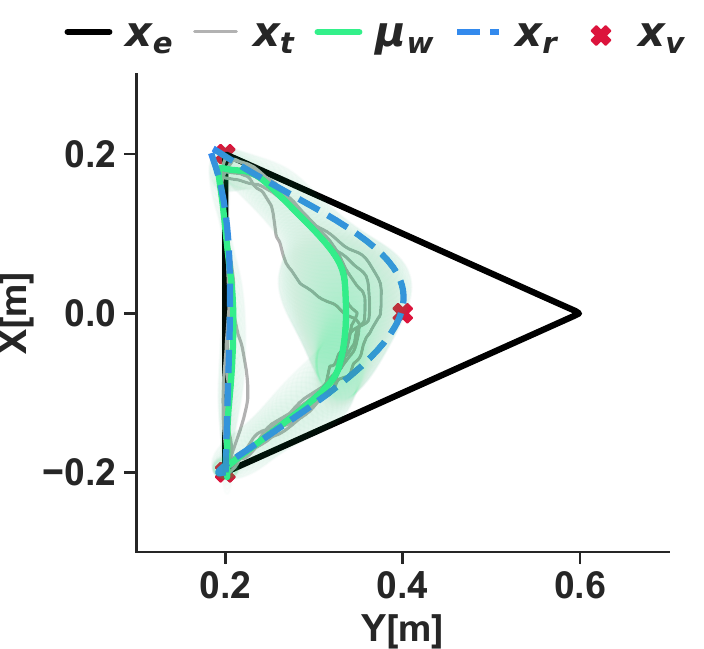}
\label{Fig4(b):path-encoding-gmm}}  
\subfigure[]{\includegraphics[width=0.23\linewidth]{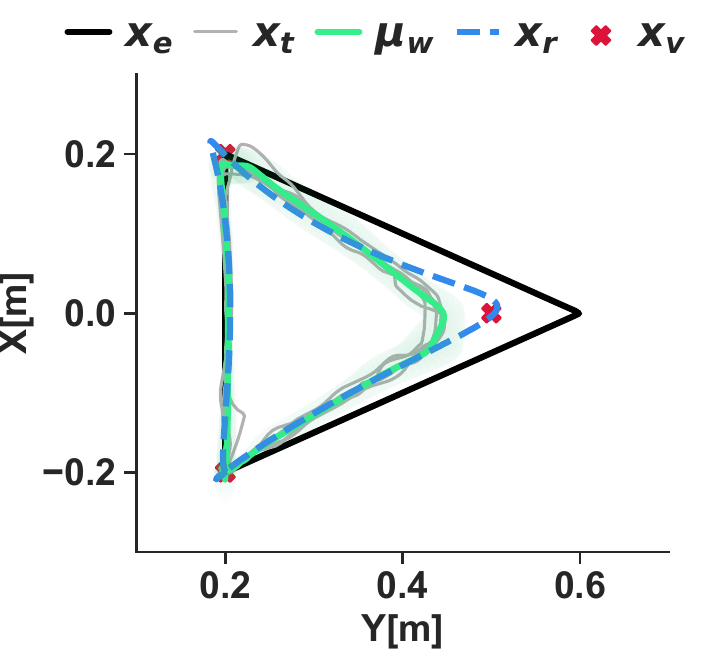}  
\label{Fig4(c):path-encoding-gmm}}   
\subfigure[]{\includegraphics[width=0.23\linewidth]{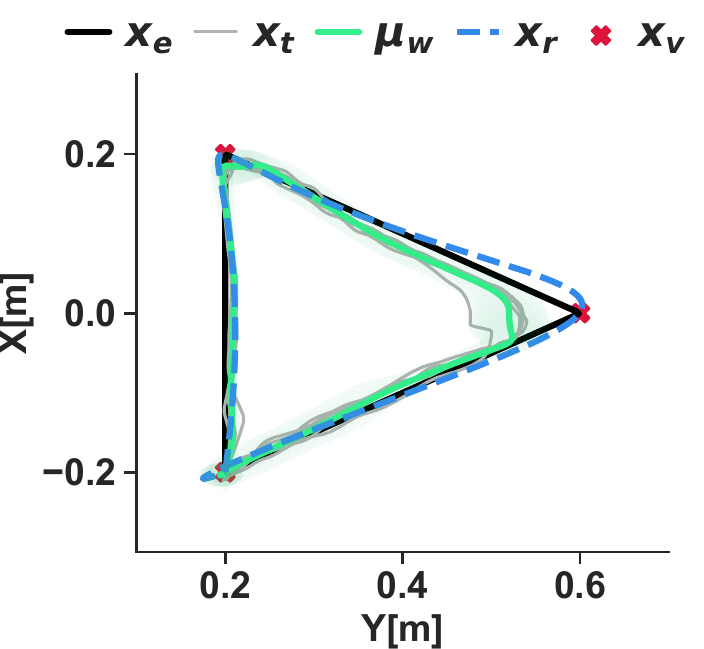}     
\label{Fig4(d):path-encoding-gmm}} 
\caption{
Reproduction of therapist-informed therapy skill for Task \#1 performed by patient at Stage \#2.~(a)~Iteration 1;~(b)~Iteration 3;~(c)~Iteration 6;~(d)~Iteration 9. 
}  
\label{Fig2:theraputic-movement-task-1-reproduction}  
\vspace{-0.6cm}  
\end{figure*}    
\begin{figure*}[!t] 
\setlength{\abovecaptionskip}{-0.15cm}   
\centering  
\subfigure[]{\includegraphics[width=0.23\linewidth]{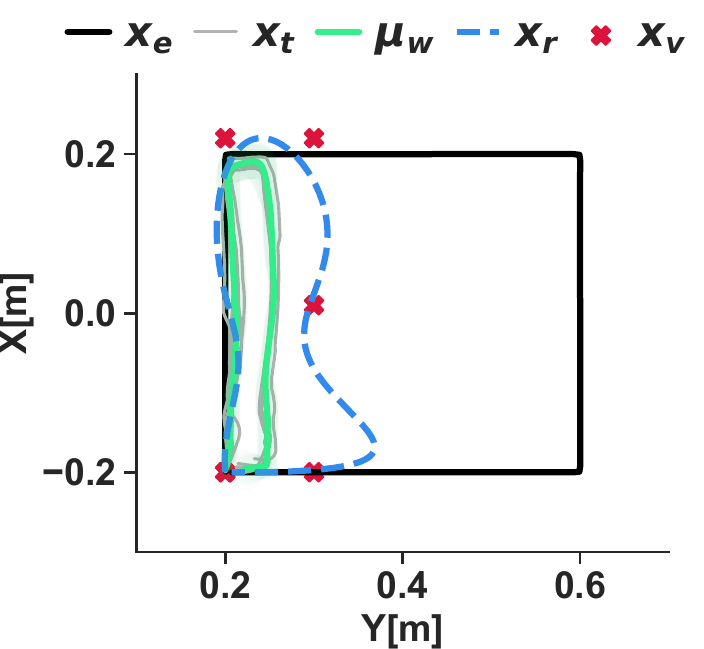}   
\label{Fig4(a):path-encoding-gmm}} 
\subfigure[]{\includegraphics[width=0.23\linewidth]{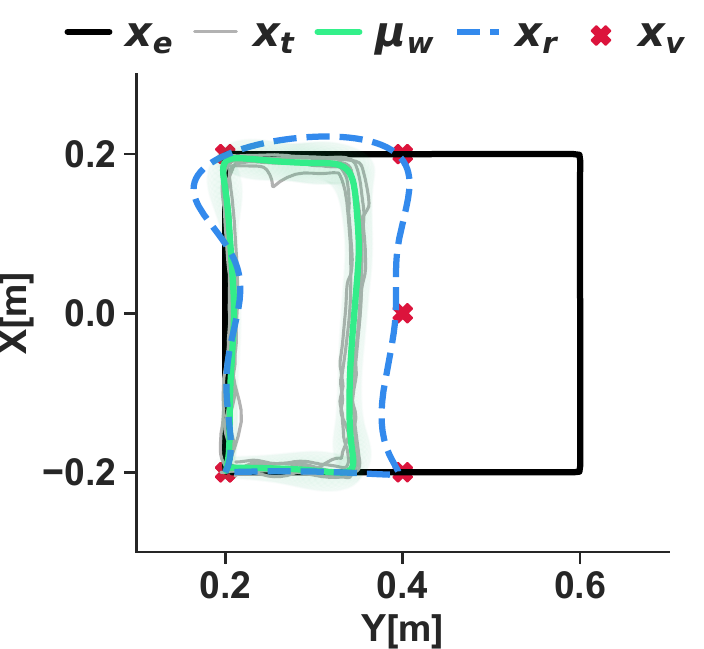}
\label{Fig4(b):path-encoding-gmm}}  
\subfigure[]{\includegraphics[width=0.23\linewidth]{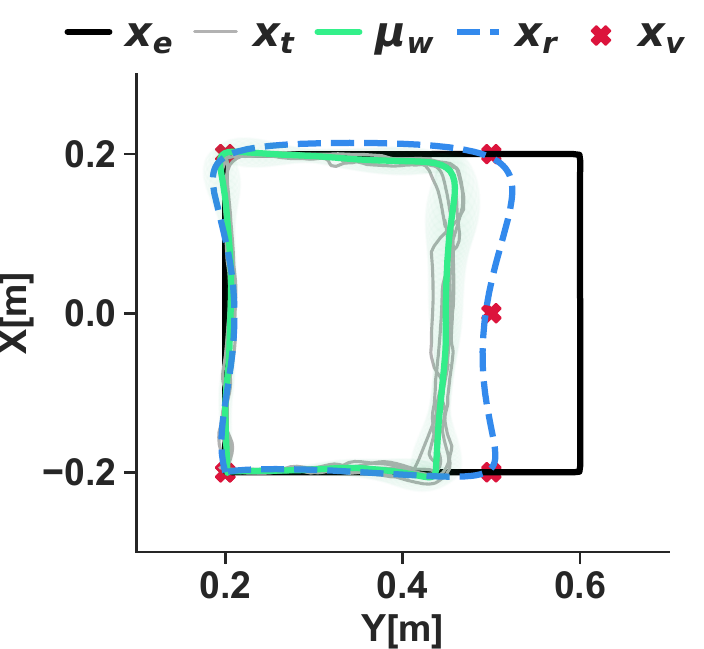}  
\label{Fig4(c):path-encoding-gmm}}   
\subfigure[]{\includegraphics[width=0.23\linewidth]{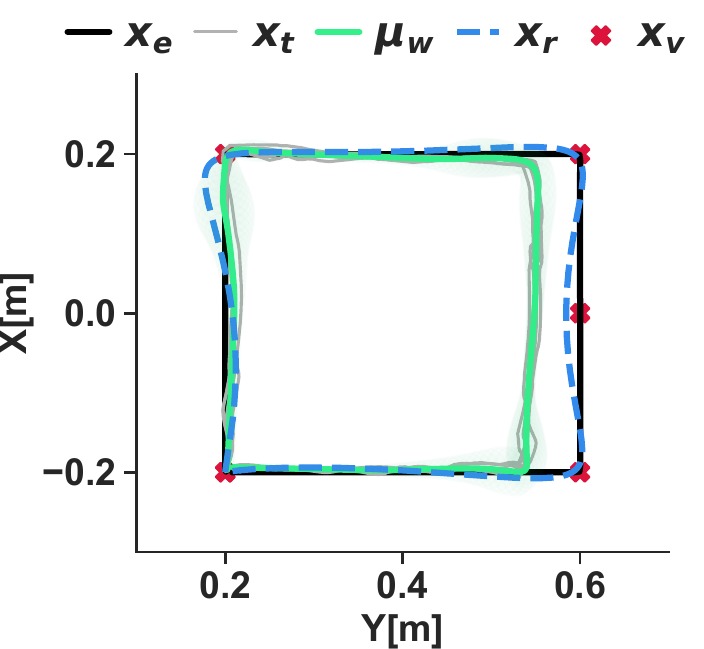}  
\label{Fig4(d):path-encoding-gmm}} 
\caption{
Reproduction of therapist-informed therapy skill for Task \#2 performed by patient at Stage \#2.~(a)~Iteration 1;~(b)~Iteration 3;~(c)~Iteration 6;~(d)~Iteration 9. 
} 
\label{Fig2:theraputic-movement-task-2-reproduction}   
\vspace{-0.7cm}  
\end{figure*}   

The desired motion of Task \#2~(a rectangular trajectory) was visualized by black lines as shown in Fig.~\ref{Fig2:theraputic-movement-task-2}. 
At $i$-th therapy iteration, $J = 5$ actual therapy trajectories, visualized by gray lines, are collected to encode the motion preferences~(green lines) using GMMs with $C_2=10$ components. 
Following the same assumption of therapist's skills in Task \#1, $L_2=5$ via-points were extracted from the therapist-informed corrective force visualized by red scatters. 
The therapist-informed corrective force plotted in the second row indicates that the therapist only informed three via-points for patient therapy to reach three corners of the rectangular trajectory. 
The reference trajectory for the next therapy iteration was then generated and is illustrated by the blue line. 
For a fair comparison, rather than defining a specific rehabilitation termination condition, $I=10$ therapy iterations were conducted to evaluate the effectiveness of Baseline \#1 and Baseline \#2 on both Task \#1 and Task \#2 performed by the patient using the same elastic band. 
Baseline \#1 was implemented using the variable impedance controller given the desired motion~(black lines in Fig.~\ref{Fig2:theraputic-movement-task-1} and Fig.~\ref{Fig2:theraputic-movement-task-2}) as the reference trajectory. The stiffness is modulated according to the tracking error and the damping matrix was derived from the stiffness using a fixed relation. 
Baseline \#2 was implemented using a zero-impedance controller for the patient-side robot to allow patient maximum participant, while directly adding the therapist-informed corrective force from therapist-side robot. 
The average values of two metrics are calculated for three therapy iterations. 
As shown in Fig.~\ref{Fig6:-metric1-task1} and \ref{Fig6:-metric1-task2}, Baseline \#1 results in the large corrective force when the patient's actual motion is far from the desired motion~(see therapy iteration 1). 
Although the Baseline \#2 can reduce the corrective force from the robot and encourage patient's active engagement, as illustrated in Fig.~\ref{Fig6:-metric2-task1} and \ref{Fig6:-metric2-task2}, movement smoothness is affected by the therapist-informed corrective force. 
In summary, the proposed controller can not only maximally encourages the patient's active participant without introducing excessive corrective force, but also ensures that the smoothness of the movement therapy remains unaffected. 
\subsubsection{Reproducing Therapist's Therapy Skill} 
The collected therapist-informed via-points in dataset $\md_T$ are employed to obtain the optimal weight matrix $\mbB^{\star}$ for Task \#1 and Task \#2 using \textit{PLSRegression} implementation from the open-source \textit{scikit-learn} package. 
Afterwards, via-points are generated by the regression function for patient at another recovery Stage \#2, which is acted by an able-bodied participant wearing another elastic band with greater elasticity~(see Fig.~\ref{fig:experimental-setups}(b)). 
Results in Fig.~\ref{Fig2:theraputic-movement-task-1-reproduction} and Fig.~\ref{Fig2:theraputic-movement-task-2-reproduction} indicate that the patient can achieve the set desired motion for both Task \#1 and Task \#2 after $I=10$ therapy iterations without additional instruction from the therapist. 
The performance of the reference trajectory deformation depends on the hyperparameters, as summarized in Table~\ref{tab-para-system-performance}. 
In this study, we assume that the therapist primarily focused on providing corrective forces to help the patient capture the key motion features with $L_1=4$ and $L_2=5$ via-points.    
\vspace{-0.2cm} 
\section{Discussion and Conclusion} 
This article proposes a novel therapist-in-the-loop AAN therapy framework. 
Its effectiveness and advantages in utilizing therapist's expertise over state-of-the-art methods were validated using a telerobotics-mediated upper-limb rehabilitation system. 
In \cite{liu2020home}, the therapist needs to re-plan the entire reference trajectory for each therapy iteration or to adapt impedance parameters for varied therapy intensity~\cite{pezeshki2023cooperative}. 
Our framework aims to maintain patient motion preferences while partially deforming the reference trajectory, thereby maximizing the active participation of patients and effectively incorporating therapist-informed corrective forces. 
Moreover, the approaches in \cite{luciani2024therapists},\cite{hou2025bilateral} rely on low-impedance robotic systems or external sensors to detect the corrective force and apply them in patient-side therapy. 
By contrast, we build a latent space to infer via-points for trajectory partial deformation directly from therapist-informed corrective force. 
The results demonstrate that our framework can ensure the safety by avoiding the large corrective force and mitigating the effect of communication delay for movement therapy. 
From a practical perspective, our framework enables therapists to provide corrective forces through the robot without the need to program reference trajectories or manually tune control parameters. 

This study provides a feasibility evaluation of the proposed controller; however, the experimental validation is limited to two able-bodied participants under simulated motion disabilities. In future work, clinical studies will be conducted with patients at different stages of rehabilitation to further assess the therapeutic benefits of the proposed approach and to analyze the associated biomechanical responses. 
In this paper, the partial least squares regression model used to reproduce therapist skills is task-dependent and relies on labeled data collected for each specific task. Future work will therefore focus on developing in-context regression functions or learning methods capable of reproducing therapeutic skills across different tasks. In addition, the proposed telerobotics-mediated system will be extended to telerehabilitation applications by integrating advanced communication protocols~(e.g., 5G).  
    
\vspace{-0.4cm} 
\bibliographystyle{IEEEtran}   
\bibliography{
IEEEabrv,
utils/reference,
utils/my_references
}   

\end{document}